\documentclass[aps,twocolumn,superscriptaddress]{revtex4}
\usepackage{bm}
\usepackage[dvipdfmx]{graphicx}
\usepackage{amsmath}
\usepackage{color}

\begin{document}
\title {Real-time dynamics of the photoinduced topological state in organic conductor $\alpha$-(BEDT-TTF)$_2$I$_3$ under continuous-wave and pulse excitations}

\author{Yasuhiro Tanaka}
\affiliation{Department of Applied Physics, Waseda University, Okubo, Shinjuku-ku, Tokyo 169-8555, Japan}
\author{Masahito Mochizuki}
\affiliation{Department of Applied Physics, Waseda University, Okubo, Shinjuku-ku, Tokyo 169-8555, Japan}
\begin{abstract}
We theoretically study the real-time dynamics of the photoinduced topological phase transition to a nonequilibrium Floquet Chern insulator in an organic conductor $\alpha$-(BEDT-TTF)$_2$I$_3$, which was recently predicted using the Floquet theory. By using a tight-binding model of $\alpha$-(BEDT-TTF)$_2$I$_3$ that hosts a pair of tilted Dirac-cone bands at the Fermi level, we solve the time-dependent Schr\"odinger equation and obtained time evolutions of physical quantities for continuous-wave and pulse excitations with circularly polarized light. We demonstrate that, for the continuous-wave excitations, time profiles of the Chern number and the Hall conductivity show indications of the Floquet topological insulator. We argue that the Hall conductivity exhibits a slow oscillation with its frequency corresponding to a photoinduced direct gap determined by the Floquet band structure. With pulse excitations, transient excitation spectra are obtained, from which we infer the formation of Floquet bands and the gap opening at the Dirac point during the pulse irradiation. This dynamical gap formation is also manifested by the slow oscillation component of the Hall conductivity; that is, its frequency increases with time toward the pulse peak at which it nearly coincides with the photoinduced direct gap. The relevance of the results to experiments is also discussed.
\end{abstract}
\maketitle

\section{Introduction}
Photoinduced changes in electronic properties of solids offer a fascinating research field in condensed matter physics~\cite{Kirilyuk_RMP10,Bukov_AP15,Ishihara_JPSJ19}. In particular, the emergence of topologically nontrivial phases under light irradiation has attracted much attention. A representative phenomenon was predicted theoretically in graphene~\cite{Oka_PRB09}, in which its massless Dirac fermions acquire a topological gap when subjected to irradiation of circularly polarized light. Such nonequilibrium states are called Floquet topological insulators, which are characterized by a nonzero Chern number and their novel light-induced anomalous Hall effect~\cite{Oka_PRB09,Kitagawa_PRB11,Kitagawa_PRB11n2}. 

For systems driven by a time-periodic external field such as continuous-wave light, the Floquet theorem provides a mapping from the original time-periodic Hamiltonian to an effective static Hamiltonian, called the Floquet Hamiltonian, that enables us to study nonequilibrium physical properties in a time-independent manner~\cite{Rahav_PRA03,Kitagawa_PRB11,Kitagawa_PRB11n2,Goldman_PRX14,Mikami_PRB16}. This approach has played a major role in studying various photoinduced phases including the Floquet topological insulator phase~\cite{Linder_NP11,Ezawa_PRL13,Grushin_PRL14,Zou_PRB16,Takasan_PRB17,Kitayama_PRR20}. However, to observe these phases experimentally, an intense electric field of light is often required; therefore, using ultrafast laser pulses is more feasible rather than a continuous-wave light. Indeed, for graphene, the light-induced anomalous Hall effect has been observed using ultrafast pulses of circularly polarized light~\cite{Mclver_NP20}. In this respect, how characteristic properties of the Floquet topological insulator phase appear under pulse excitations, especially in transient dynamics that are not accessible with approaches based on the Floquet theorem, is of great interest. Theoretical investigations in this direction have been reported recently, which was based on the time-dependent Schr\"odinger equation ~\cite{Sentef_NatCom15,Schuler_PRB17,Gavensky_PRB18,Sato_NJP19,Sato_PRB19,Schuler_PRX20,Aschlimann_Arxiv21}. 

\begin{figure}
\includegraphics[scale=0.25]{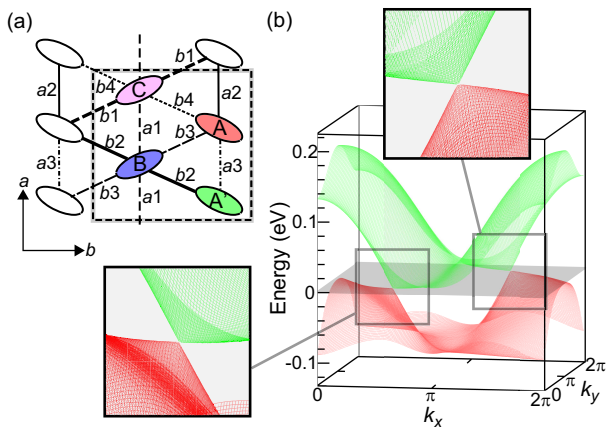}
\caption{(a) Schematic illustration of the conductive BEDT-TTF layer in $\alpha$-(BEDT-TTF)$_2$I$_3$. The dashed rectangle marks the unit cell, which contains four molecular sites (A, A$^{\prime}$, B, C). (b) Energy dispersion relations for the two highest bands ($\nu=3$, $4$), in which the tilted Dirac cones are located at the Fermi level that is taken as 0 eV. Enlarged views of the tilted Dirac cones are also shown.}
\label{fig:fig1}
\end{figure}

To date, direct evidence of a photoinduced topological state has been observed in surface Dirac fermions of a topological insulator Bi$_2$Se$_3$~\cite{Wang_SCI13}, for which the time- and angle-resolved photoemission spectroscopy measurements showed a gap opening at the surface Dirac point by irradiation with circularly polarized light. However, the number of such materials has been severely limited. Moreover, the light-induced anomalous Hall effect, which is another approach to assess photoinduced topological changes, has been observed only in graphene~\cite{Mclver_NP20}. The search for new candidate materials hosting the Floquet topological insulator phase is thus indispensable to deepen our understanding of this novel phenomenon. One material of interest for this purpose is the quasi-two-dimensional organic conductor $\alpha$-(BEDT-TTF)$_2$I$_3$. Here BEDT-TTF stands for bis(ethylenedithio)tetrathiafulvalene. At ambient pressure, this compound exhibits a charge-order insulating state below 135 K~\cite{Bender_MCLC84,Kakiuchi_JPSJ07,Kino_JPSJ95,Seo_JPSJ00}. Applying pressure destabilizes the charge order, and a zero-gap state with tilted Dirac cones at the Fermi level appears (Fig. \ref{fig:fig1})~\cite{Katayama_JPSJ06,Katayama_EPJB08,Tajima_JPSJ06}. The appearance of this zero-gap state under pressure was verified by theoretical studies based on the first-principles calculations~\cite{Kino_JPSJ06} and supported experimentally by transport measurements~\cite{Tajima_JPSJ06,Kajita_JPSJ14}. A recent theoretical study based on the Floquet theory~\cite{Kitayama_PRR20} has shown that this zero-gap state turns into the Floquet topological insulator phase by irradiation with circularly polarized light, and that rich phase diagrams are obtained in the plane of the frequency and amplitude of light electric field. This motivates our study of the transient dynamics associated with the Floquet topological insulator phase in this compound. 

In this paper, we investigate the photoinduced topological state and the formation of Floquet-dressed bands in $\alpha$-(BEDT-TTF)$_2$I$_3$ from the viewpoint of real-time dynamics. By numerically solving the time-dependent Schr\"odinger equation for a tight-binding model of $\alpha$-(BEDT-TTF)$_2$I$_3$ with ac electric field of circularly polarized light, we obtain photoinduced dynamics for continuous-wave and pulse excitations. We first examine the continuous-wave excitations and calculate time profiles of the Chern number and the Hall conductivity, from which we explore new physics of the photoinduced topological phase transitions and the resulting Floquet Chern insulator phase. We also discuss what aspects of these photoinduced phenomena and nonequilibrium phases the Floquet theory captures or misses. Also, we show that the Hall conductivity has an oscillation component with frequency much smaller than the light frequency. The center of the oscillation coincides with the quantized Hall conductivity characterized by a nonzero Chern number, whereas its frequency corresponds to the direct gap extracted from the Floquet bands. We then consider the pulse excitations and calculate the transient excitation spectra and the Hall conductivity. These quantities reveal the Floquet band formation and dynamical gap opening in a time-resolved manner, which are not found from the Floquet theory. We discuss the relevance of our results to experiments in $\alpha$-(BEDT-TTF)$_2$I$_3$. 

\section{Model and Method}
We consider the two-dimensional tight-binding model for $\alpha$-(BEDT-TTF)$_2$I$_3$ [Fig. \ref{fig:fig1}(a)] defined by~\cite{Kitayama_PRR20,Tanaka_JPSJ10,Miyashita_JPSJ10}
\begin{equation}
{\mathcal H}(\tau)=\sum_{\langle ij\rangle,\sigma}t_{i,j}e^{i(e/\hbar){\bm \delta}_{i,j}\cdot {\bm A}(\tau)}c^{\dagger}_{i\sigma}c_{j\sigma}+{\rm H.c.},
\label{eq:ham}
\end{equation}
where $\langle ij\rangle$ signifies the summation over pairs of nearest-neighbor sites, and $c^{\dagger}_{i\sigma}$ ($c_{i\sigma}$) denotes the creation (annihilation) operator for an electron with spin $\sigma$ at the $i$th site. The effects of light electric field are incorporated by the Peierls phase for the transfer integrals $t_{i,j}$, through which the Hamiltonian depends on time $\tau$. We define ${\bm \delta}_{i,j}={\bm r}_j-{\bm r}_i$ with ${\bm r}_i$ being the position vector for the $i$th site. We note that the model in Eq. (\ref{eq:ham}) describes the energy bands of $\alpha$-(BEDT-TTF)$_2$I$_3$ near the Fermi level and these bands are constructed from a single orbital, the highest occupied molecular orbital of a BEDT-TTF molecule \cite{Kino_JPSJ06}. Since the bands originating from other molecular orbitals are energetically separated from these bands, the dipole transition terms do not appear in Eq. (\ref{eq:ham}). For the global band structure of this compound, see Ref.~\cite{Kino_JPSJ06}. Hereafter, we use natural units with $e=\hbar=1$. The lattice constant along the $a$ axis is chosen as the unit of length. Note that the unit cell containing four molecules A, A$^{\prime}$, B, and C [Fig. \ref{fig:fig1}(a)] is rectangular with different lattice constants of $a$ and $b$ in reality. However, because the difference is small, we assume a squared unit cell in the present study. We have confirmed that this simplification does not alter the results and conclusions so much. The vector potential for the ac electric field of circularly polarized light is given by ${\bm A}(\tau)$. For the continuous-wave excitations, we write
\begin{equation}
{\bm A}(\tau)=A_0f(\tau)(\cos \omega \tau, \sin \omega \tau), 
\end{equation}
where $A_0=E^{\omega}/\omega$ with $E^{\omega}$ and $\omega$ being, respectively, the amplitude and frequency of the ac electric field of light. We consider a left-handed circularly polarized light unless otherwise specified. A factor $f(\tau)$ defined by $f(\tau)=e^{-\tau^2/\tau_{\rm ac}^2}$ for $\tau\leq0$ and $f(\tau)=1$ for $\tau>0$ is set to obtain a quasiadiabatic time evolution~\cite{Dalessio_NatCom15,Mochizuki_APL18}. We use $\tau_{\rm ac}=300\,T$, where $T=2\pi/\omega$ is the period of light. In contrast, for the pulse excitations, we use
\begin{equation}
{\bm A}(\tau)=A_0\exp \Big[-\frac{(\tau-\tau_{\rm pu})^2}{2\sigma^2_{\rm pu}}\Big] (\cos \omega \tau, \sin \omega \tau), 
\end{equation}
which gives a pulse of width $\sigma_{\rm pu}$ centered around time $\tau_{\rm pu}$. The time evolution of the system is calculated by the time-dependent Schr\"odinger equation
\begin{equation}
|\psi_{{\bm k},\nu}(\tau+d\tau)\rangle={\mathcal T}{\rm exp}
\Bigl[-i\int^{\tau+d\tau}_{\tau} d\tau^{\prime}{\mathcal H}_{{\bm k}}(\tau^{\prime})
\Bigr]|\psi_{{\bm k},\nu}(\tau)\rangle,
\end{equation}
where ${\mathcal H}_{\bm k}$ denotes the momentum representation of the Hamiltonian matrix in Eq. (\ref{eq:ham}), $|\psi_{{\bm k},\nu}(\tau)\rangle$ the $\nu$th ($\nu=1$--$4$) one-particle state with wave vector ${\bm k}$ at time $\tau$, and ${\mathcal T}$ the time-ordering operator. The spin index $\sigma$ is omitted for brevity. This equation is numerically solved by writing~\cite{Kuwabara_JPSJ95,Terai_PTPS93,Tanaka_JPSJ10}
\begin{equation}
|\psi_{{\bm k},\nu}(\tau+\Delta\tau)\rangle
\simeq {\rm exp}\Bigl[-i\Delta \tau {\mathcal H}_{{\bm k}}(\tau+\Delta \tau/2)\Bigr]
|\psi_{{\bm k},\nu}(\tau) \rangle,
\end{equation}
which gives the time-evolving one-particle states within an error of the order of $(\Delta \tau)^3$. We use $\Delta \tau=0.01$ throughout, which guarantees sufficient numerical accuracy. For the values of $t_{i,j}$, we adopt those for a compound under uniaxial pressure of $P=4$ kbar. They are estimated from the relation $t_{i,j}=t_{i,j}^{\rm ap}(1+K_{i,j}P)$ where the coefficients $K_{i,j}$ are given in Ref.~\cite{Kobayashi_JPSJ04}. The transfer integrals at ambient pressure $t_l^{\rm ap}$ are given by $t_{b1}^{\rm ap}=0.127$, $t_{b2}^{\rm ap}=0.145$, $t_{b3}^{\rm ap}=0.062$, $t_{b4}^{\rm ap}=0.025$, $t_{a1}^{\rm ap}=-0.035$, $t_{a2}^{\rm ap}=-0.046$, and $t_{a3}^{\rm ap}=0.018$~\cite{Kakiuchi_JPSJ07}; all values are in units of eV. Here, $l$ is the index that specifies the bonds [see Fig. \ref{fig:fig1}(a)]. We take $L_a=L_b=200$ with $L_a$ ($L_b$) being the number of unit cells in the $a$ direction ($b$ direction). The energy dispersion relations of the two highest bands ($\nu=3, 4$) are given in Fig. \ref{fig:fig1}(b). The electron density of this compound is specified at 3/4 filling. Before the photoexcitation, the system is a zero-gap state with the Fermi level coinciding with the contact points of the tilted Dirac cones, which are located at ${\bm k}^{\pm}/\pi=(1,1)\pm (0.40, 0.67)$.

\section{Continuous-wave excitations}
\begin{figure}
\includegraphics[scale=1.0]{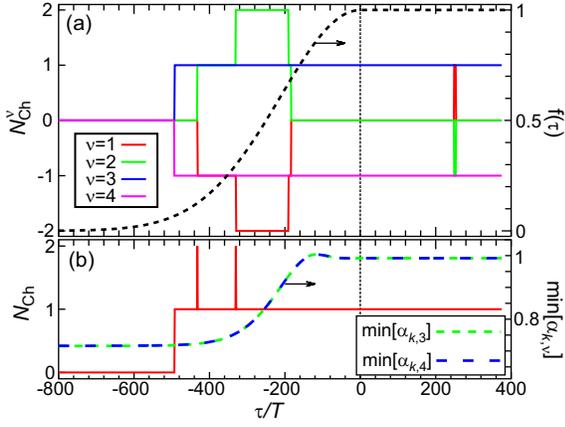}
\caption{(a) Time profiles of the Chern numbers $N_{\rm Ch}^{\nu}$, for which $\nu$ is the band index. The time dependence of $f(\tau)$ is also depicted. (b) Time profiles of $N_{\rm Ch}$ and ${\rm min}[\alpha_{{\bm k},\nu}]$ with $\nu=3$ and 4. We use continuous-wave excitations with $\omega=0.8$ and $A_0=0.8$. The results are shown at stroboscopic times.}
\label{fig:fig2}
\end{figure}

We next consider continuous-wave excitations. The light frequency is chosen at $\omega=0.8$ (in units of eV) for which the Floquet theory predicts the emergence of a Floquet topological insulator phase for nonzero $A_0$~\cite{Kitayama_PRR20}. In Fig. \ref{fig:fig2}(a), we plot the time profile of Chern numbers $N_{\rm Ch}^{\nu}(\tau)$. Following the computational method to calculate the Chern number in equilibrium proposed in Ref.~\cite{Fukui_JPSJ05}, we write $N_{\rm Ch}^{\nu}(\tau)$ in nonequilibrium as
\begin{equation}
N_{\rm Ch}^{\nu}(\tau)=\frac{1}{2\pi i}\sum_{\bm k}F^{\nu}({\bm k},\tau),
\end{equation}
where
\begin{equation}
\begin{split}
F^{\nu}({\bm k},\tau)&=\ln [U^{\nu}_x({\bm k},\tau)U^{\nu}_y({\bm k}+{\bm \delta}_x,\tau) \\
&\times U^{\nu}_x({\bm k}+{\bm \delta}_y,\tau)^{-1}U^{\nu}_y({\bm k},\tau)^{-1}], 
\label{eq:F}
\end{split}
\end{equation}
which is defined by the principal value of the logarithm. 
In Eq. (\ref{eq:F}), we define 
\begin{equation}
{\bm \delta}_{x}=(2\pi/L_b,0),\  {\bm \delta}_{y}=(0,2\pi/L_a),
\end{equation}
and 
\begin{equation}
U^{\nu}_{\mu}({\bm k})=\langle \psi_{{\bm k},\nu}(\tau)|\psi_{{\bm k}+{\bm \delta}_{\mu},\nu}(\tau)\rangle/|\langle \psi_{{\bm k},\nu}(\tau)|\psi_{{\bm k}+{\bm \delta}_{\mu},\nu}(\tau)\rangle|, 
\end{equation}
with $\mu=x,y$. The results for $A_0=0.8$ are shown at stroboscopic times $\tau/T=n$ with $n$ integer. We note that for $\alpha$-(BEDT-TTF)$_2$I$_3$, the contact points come from the accidental degeneracy of the energy bands in the thermodynamic limit~\cite{Herring_PR37,Suzumura_JPSJ16}. This is in contrast to graphene, for which they are exactly on the symmetric points in the Brillouin zone (BZ). Because of this property, for finite size systems, the gap closing points between the two bands $\nu=3$ and $4$ are absent in the discrete BZ. This allows temporal changes of the Chern numbers of these bands reflecting a topological phase transition in our simulation~\cite{Dalessio_NatCom15,Ge_PRA17}. Before photoexcitation, we have $N^{\nu}_{\rm Ch}=0$ for all $\nu$; the initial state is topologically trivial. Although the Chern numbers of the two lowest bands, $N^{1}_{\rm Ch}$ and $N^{2}_{\rm Ch}$, exhibit a complex time dependence for $-440\gtrsim \tau/T\gtrsim -180$ for which the temporal variation in $f(\tau)$ becomes large, they are basically conserved at zero for $\tau/T> -180$. We note that the electric field is maximal for $\tau\geq 0$. The value of $N^{3}_{\rm Ch}$ ($N^{4}_{\rm Ch}$) changes from 0 to 1 ($-1$) at $\tau/T\sim -500$ and thereafter remains unchanged. In Fig. \ref{fig:fig2}(b), we plot the time profile of $N_{\rm Ch}(\tau)=\sum_{\nu=1}^3 N_{\rm Ch}^{\nu}(\tau)$ at stroboscopic times demonstrating the appearance of a topologically nontrivial state with $N_{\rm Ch}=1$. This result is consistent with that obtained by the Floquet theory~\cite{Kitayama_PRR20}. Here we mention that in the time profile of $N_{\rm Ch}$ there appear two peaks at $\tau/T\sim -430$ and $-330$. They come from a slight difference between the stroboscopic times at which $N^{1}_{\rm Ch}$ and $N^{2}_{\rm Ch}$ change [Fig. \ref{fig:fig2}(a)]. Except for these two peaks, we have $N^{1}_{\rm Ch}=-N^{2}_{\rm Ch}$ for all stroboscopic times and thus $N_{\rm Ch}=N^{3}_{\rm Ch}$ holds. To verify the dynamically prepared Floquet topological insulator phase in more detail, we calculate the overlap $\alpha_{{\bm k},\nu}$ which is defined by
\begin{equation}
\alpha_{{\bm k},\nu}=|\langle \psi^F_{{\bm k},\nu}|\psi_{{\bm k},\nu}(\tau)\rangle|, 
\end{equation}
where $|\psi^F_{{\bm k},\nu}\rangle$ denotes the ground-state wave function of the Floquet Hamiltonian ${\mathcal H}_F$ for a system subject to a continuous ac electric field with amplitude $E^{\omega}$. Because we have~\cite{Dalessio_NatCom15}
\begin{equation}
U(T,0)=e^{-i{\mathcal H}_FT},
\end{equation}
where $U(T,0)$ denotes the time-evolution operator over one period, we obtain $|\psi^F_{{\bm k},\nu}\rangle$ by diagonalizing $U(T,0)$. For $\omega=0.8$, the off-resonance condition is satisfied where the Floquet bands with different photon numbers do not overlap, making it possible to identify the $\nu$th Floquet band unambiguously~\cite{Kitayama_PRR20,Dalessio_NatCom15}. In Fig. \ref{fig:fig2}(b), we show the time evolutions of ${\rm min}[\alpha_{{\bm k},3}]$ and ${\rm min}[\alpha_{{\bm k},4}]$, for which ${\rm min}[\alpha_{{\bm k},\nu}]$ is the minimum of $\alpha_{{\bm k},\nu}$ in the BZ for a fixed $\nu$. They are nearly identical to each other; ${\rm min}[\alpha_{{\bm k},3}]$ and ${\rm min}[\alpha_{{\bm k},4}]$ show a gradual increase in accordance with $f(\tau)$ and then approach $1$ at $\tau/T=0$. The overlaps $\alpha_{{\bm k},3}$ and $\alpha_{{\bm k},4}$ have their minimum values at the contact points where the photoinduced gap opens. We note that for $\tau/T>0$, $\alpha_{{\bm k},\nu}$ is conserved~\cite{Dalessio_NatCom15}. These results indicate that the time-evolving one-particle states are well described by the ground state of ${\mathcal H}_F$. 

\begin{figure}
\includegraphics[scale=1.0]{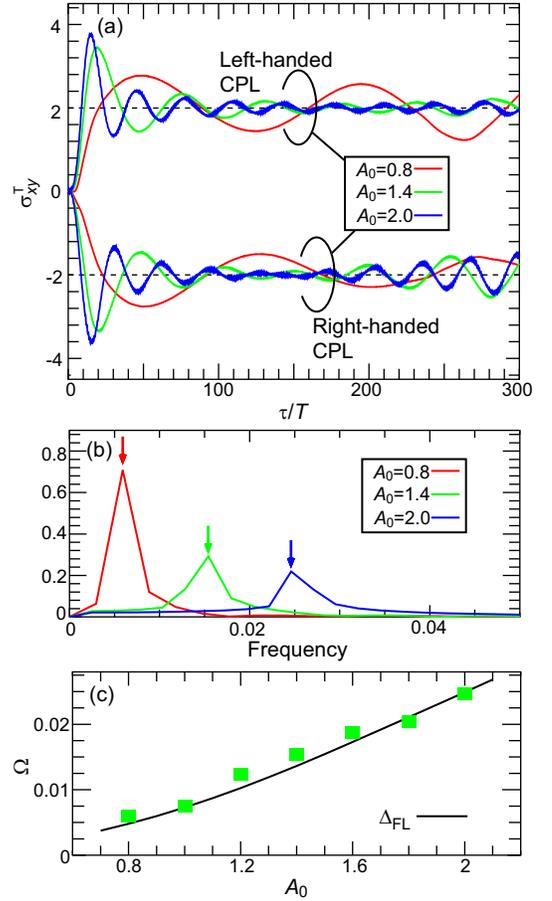}
\caption{(a) Time evolution of $\sigma^T_{xy}$ for different values of $A_0$ with $\omega=0.8$ and $E_{\rm dc}^y=2\times 10^{-5}$. The results with left-handed (L) and right-handed (R) circularly polarized light are shown. The dashed horizontal lines indicate the values of the quantized Hall conductivity for Floquet topological insulators with $N_{\rm Ch}=\pm 1$. (b) Fourier transform of $\sigma^T_{xy}$. The peak position in each spectrum is indicated by an arrow. (b) Peak frequency $\Omega$ as a function of $A_0$, where the solid line marks the magnitude of the direct gap calculated from the Floquet bands with $\nu=3$ and $4$.}
\label{fig:fig3}
\end{figure}

In regard to the physical quantity that characterizes the Floquet topological insulator, we calculate the Hall conductivity $\sigma_{xy}$. For this purpose, we consider the vector potential ${\bm A}_{\rm dc}$ for a static electric field ${\bm E}_{\rm dc}=(E_{\rm dc}^x,E_{\rm dc}^y)$ that is switched on at $\tau=0$; we have
\begin{equation}
{\bm A}_{\rm dc}(\tau)=-\gamma(\tau)(E^x_{\rm dc}\tau,E^y_{\rm dc}\tau), 
\end{equation}
where we introduce a factor $\gamma(\tau)$ that is given by $\gamma(\tau)=0$ for $\tau<0$ and $\gamma(\tau)=1-e^{-\tau^2/\tau^2_{\rm dc}}$ for $\tau\geq0$ with $\tau_{\rm dc}=10T$. We define the current operator by
\begin{equation}
{\bm J}=-\frac{\partial {\mathcal H}({\bm A}+{\bm A}_{\rm dc})}
{\partial {\bm A}_{\rm dc}}, 
\end{equation}
from which we obtain $\sigma_{xy}$,
\begin{equation}
\sigma_{xy}=\frac{1}{2\pi N}\frac{\langle J_x \rangle}{E_{\rm dc}^y},
\end{equation}
where $N=L_aL_b$ is the total number of unit cells. We set $E_{\rm dc}^x=0$ and $E_{\rm dc}^y\ll E^{\omega}$ so that the static electric field does not affect the photoinduced dynamics qualitatively. Because the obtained $\sigma_{xy}$ strongly oscillates with the light frequency $\omega$, we compute a time-averaged quantity $\sigma_{xy}^T$ over one period,
\begin{equation}
\sigma_{xy}^T(\tau)=\int_{\tau-T/2}^{\tau+T/2}\sigma_{xy}(\tau^{\prime})d\tau^{\prime}.
\end{equation}
In Fig. \ref{fig:fig3}(a), we present the time evolution of $\sigma_{xy}^T$ for different values of $A_0$ with $\omega=0.8$ and $E_{\rm dc}^y=2\times 10^{-5}$; the results pertaining to right-handed circularly polarized light, for which we use ${\bm A}(\tau)=A_0f(\tau)(\cos \omega \tau, -\sin \omega \tau)$, are also depicted. It is apparent that $\sigma_{xy}^T$ with left-handed (right-handed) circularly polarized light exhibits an oscillation, the center of which is $2$ ($-2$) corresponding to the quantized Hall conductivity of the Floquet topological insulator with $N_{\rm Ch}=1$~\cite{Thouless_PRL82}. The frequency of the oscillation in $\sigma_{xy}^T$ is much smaller than $\omega$. To analyze this slow oscillation quantitatively, we show the Fourier transform of $\sigma_{xy}^T$ in Fig. \ref{fig:fig3}(b) for left-handed circularly polarized light. In each spectrum, there is a sharp peak, the frequency of which is denoted $\Omega$ and increases with increasing $A_0$. In Fig. \ref{fig:fig3}(c), we plot $\Omega$ as a function of $A_0$; the magnitude of the photoinduced direct gap $\Delta_{\rm FL}$ that is obtained from the eigenvalues of ${\mathcal H}_F$ is also shown. Evidently, $\Omega\sim \Delta_{\rm FL}$ holds; the slow oscillation in $\sigma_{xy}^T$ reflects the direct gap that emerges in the Floquet band structure. In the time profile of $\sigma_{xy}^T$ shown in Fig. \ref{fig:fig3}(a), we observe that the amplitude of the slow oscillation first decreases and then increases with time, which is prominent for the right-handed circularly polarized light with $A_0=1.4$ and $2.0$. We have confirmed that this behavior is not a finite-size effect. Instead, it is a beat of two oscillations with slightly different frequencies close to $\Delta_{\rm FL}$. In calculating the Hall conductivity in the photoirradiated $\alpha$-(BEDT-TTF)$_2$I$_3$, the static electric field $E^y_{\rm dc}$ is applied, which gives rise to slightly different gap amplitudes between the two Dirac points because of the tilting of the Dirac cones. The difference in the gap amplitude is estimated to be less than 10\% of the photoinduced gap.

\begin{figure*}
\includegraphics[scale=1.0]{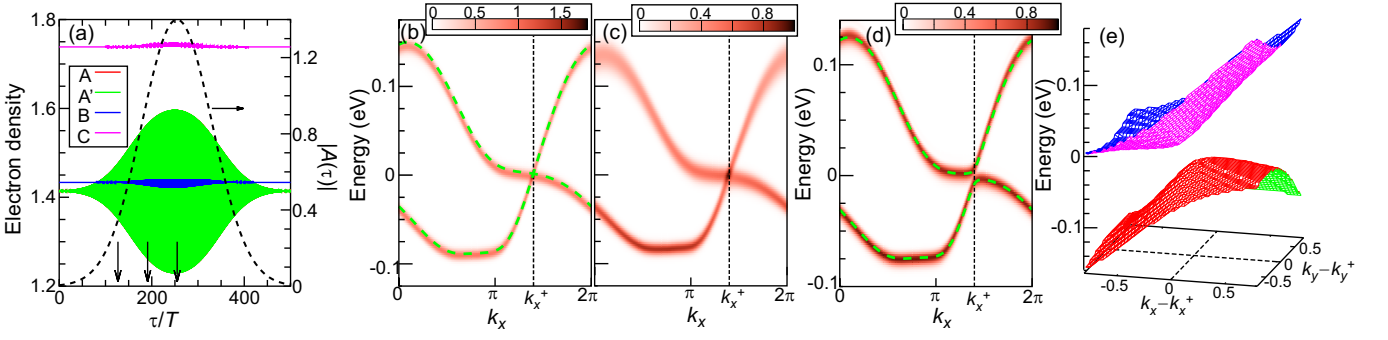}
\caption{(a) Time profiles of electron densities $n_{\alpha}$ ($\alpha=$A, A$^{\prime}$, B, C). The dashed curve plots the $\tau$ dependence of $|{\bm A}(\tau)|$. (b)--(d) Transient excitation spectra $A_{\bm k}(\varepsilon, \tau_{\rm pr})$ at $k_y=k_y^+$ as a function of $k_x$. We show results for $\tau_{\rm pr}=\tau_{\rm pu}/2$, $3\tau_{\rm pu}/4$, and $\tau_{\rm pu}$, for which the corresponding values of $\tau_{\rm pr}/T$ are indicated by the vertical arrows in (a). The dashed vertical line in each panel indicates $k_x=k_x^+$. In (b), the dashed green curve plots the energy dispersion before photoexcitation, whereas in (d) it indicates the Floquet bands obtained by diagonalizing ${\mathcal H}_F$. (e) Transient energy bands for $\tau_{\rm pr}=\tau_{\rm pu}$ near ${\bm k}={\bm k}^+$. We use $\omega=0.8$, $A_0=1.4$, $\sigma_{\rm pu}/T=76$, and $\tau_{\rm pu}/T=255$.}
\label{fig:fig4}
\end{figure*}

\section{Pulse excitations}
Next, we consider pulse excitations. For the pump pulse, we use $\omega=0.8$, $\sigma_{\rm pu}=600$ ($\sigma_{\rm pu}/T=76$), and $\tau_{\rm pu}=2\times 10^3$ ($\tau_{\rm pu}/T=255$). In Fig. \ref{fig:fig4}(a), we plot the time profiles of charge densities $n_{\alpha}$ ($\alpha=$A, A$^{\prime}$, B, C) for $A_0=1.4$. The quantities $n_{\rm A}$ and $n_{{\rm A}^{\prime}}$, which are equivalent to each other in the absence of light electric field, strongly oscillate in opposite phase. The temporal variations in $n_{\rm B}$ and $n_{\rm C}$ are small compared with those in $n_{\rm A}$ and $n_{{\rm A}^{\prime}}$. The time dependence of the oscillation amplitudes of the electron densities can be understood from the pulse shape ($|{\bm A}(\tau)|$), also shown in Fig. \ref{fig:fig4}(a).

To reveal the real-time dynamics of the Floquet band formation under pulse excitations, we calculate the transient excitation spectrum~\cite{Sentef_NatCom15,Freericks_PRL09} using
\begin{eqnarray}
A_{\bm k}(\varepsilon, \tau_{\rm pr})&=&{\rm Im} \sum_{\alpha}
\int d\tau_1d\tau_2s(\tau_1-\tau_{\rm pr}) s(\tau_2-\tau_{\rm pr}) \nonumber \\
&&\times e^{i\varepsilon(\tau_1-\tau_2)}[G^{<}_{{\bm k},\alpha\alpha}(\tau_1,\tau_2)
-G^{>}_{{\bm k},\alpha\alpha}(\tau_1,\tau_2)],
\label{eq:ak}
\end{eqnarray}
where $G^{<}_{{\bm k},\alpha\beta}(\tau_1,\tau_2)=i\langle c^{\dagger}_{{\bm k},\beta}(\tau_2)c_{{\bm k},\alpha}(\tau_1)\rangle$ and $G^{>}_{{\bm k},\alpha\beta}(\tau_1,\tau_2)=-i\langle c_{{\bm k},\alpha}(\tau_1)c^{\dagger}_{{\bm k},\beta}(\tau_2)\rangle$ denote the lesser and greater Green's functions, respectively, and $s(\tau-\tau_{\rm pr})=\frac{1}{\sigma_{\rm pr}\sqrt{2\pi}}\exp[-\frac{(\tau-\tau_{\rm pr})^2}{2\sigma_{\rm pr}^2}]$ denotes the Gaussian function for a probe pulse of width $\sigma_{\rm pr}$ centered around time $\tau_{\rm pr}$. We define the operator $c^{\dagger}_{{\bm k},\alpha}$ ($c_{{\bm k},\alpha}$) using the Fourier transform of $c^{\dagger}_{{\gamma},\alpha}$ ($c_{{\gamma},\alpha}$) where $\gamma$ indexes the unit cells. In Figs. \ref{fig:fig4}(b)--\ref{fig:fig4}(d), we present $A_{\bm k}(\varepsilon, \tau_{\rm pr})$ as a function of $k_x$ with $k_y=k_y^+$ for different values of $\tau_{\rm pr}$. We use $A_0=1.4$ and $\sigma_{\rm pr}=200$ ($\sigma_{\rm pr}/T=25$). When $\tau_{\rm pr}=\tau_{\rm pu}/2$, the structure of $A_{\bm k}(\varepsilon, \tau_{\rm pr})$ is almost identical to the energy bands in the ground state [Fig. \ref{fig:fig4}(b)]. For $\tau_{\rm pr}=3\tau_{\rm pu}/4$, $A_{\bm k}(\varepsilon, \tau_{\rm pr})$ becomes strongly blurred and at this stage the opening of the gap at the Dirac point is not visible [Fig. \ref{fig:fig4}(c)]. However, at $\tau_{\rm pr}=\tau_{\rm pu}$ for which the electric field amplitude of light has its maximum, $A_{\bm k}(\varepsilon, \tau_{\rm pr})$ exhibits a sharp structure again and a gap appears at $k_y=k_y^+$ [see Fig. \ref{fig:fig4}(d)]. The structure of $A_{\bm k}(\varepsilon, \tau_{\rm pr})$ notably coincides with the Floquet bands, which are calculated by diagonalizing ${\mathcal H}_{\rm F}$. In Fig. \ref{fig:fig4}(e), we plot the peak positions of $A_{\bm k}(\varepsilon, \tau_{\rm pr})$ near ${\bm k}={\bm k}^+$, which evidently shows that transient energy bands acquire a photoinduced gap during the pulse. 

\begin{figure}
\includegraphics[scale=1.0]{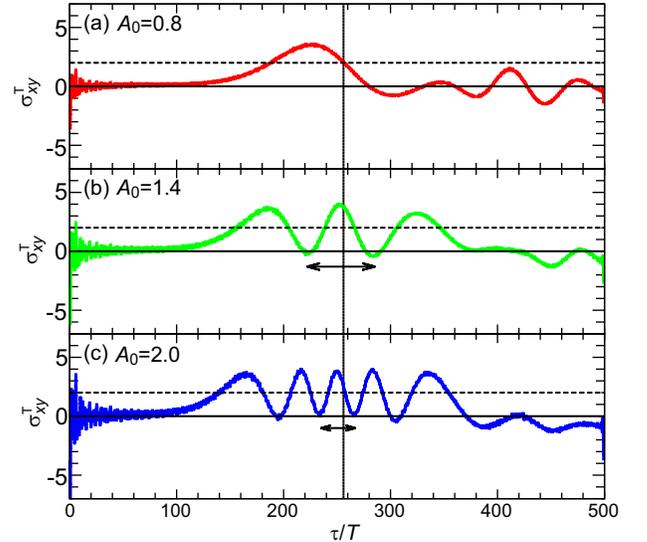}
\caption{Time profile of $\sigma^T_{xy}$ under pulse excitations for (a) $A_0=0.8$, (b) $A_0=1.4$, and (c) $A_0=2.0$. We use $\omega=0.8$ and $E_{\rm dc}^y=4\times 10^{-5}$. In each panel, the dashed horizontal line marks the value of the quantized Hall conductivity with $N_{\rm Ch}=1$, whereas the vertical line marks $\tau/T=\tau_{\rm pu}/T$ where the electric field amplitude of light becomes maximum. The double-headed arrows in (b) and (c) indicate one cycle of the slow oscillation in $\sigma^T_{xy}$ near $\tau=\tau_{\rm pu}$.}
\label{fig:fig5}
\end{figure}

In Fig. \ref{fig:fig5}, we plot $\sigma_{xy}^T$ for different values of $A_0$ with $\omega=0.8$ and $E^y_{\rm dc}=4\times 10^{-5}$. We note that a low-pass filter is used to eliminate fast oscillations in $\sigma_{xy}^T$ with frequencies around $\omega$ that originate from the pulse of circularly polarized light. Regardless of the values of $A_0$, the time profile of $\sigma_{xy}^T$ shows several features. For $\tau/T\lesssim 100$, we have $\sigma_{xy}^T\sim 0$ except for some oscillations near $\tau=0$. For $\tau/T\gtrsim 100$, $\sigma_{xy}^T$ starts to increase and oscillate around the value of the quantized Hall conductivity with $N_{\rm Ch}=1$. This oscillation is robust near the peak of the pump pulse ($\tau\sim \tau_{\rm pu}$), especially for large $A_0$, and its frequency is much smaller than $\omega$ as in the case of the continuous-wave excitations. Then, the center of this slow oscillation moves to zero at $\tau/T\sim 400$. Near $\tau=\tau_{\rm pu}$, the oscillation periods for $A_0=1.4$ and $2.0$ [Figs. \ref{fig:fig5}(b) and \ref{fig:fig5}(c)] correspond to frequencies 0.0129 and 0.0250, respectively. They are close to $\Delta_{\rm FL}$ marked in Fig. \ref{fig:fig3}(c), where we have $\Delta_{\rm FL}=0.0136$ for $A_0=1.4$ and $\Delta_{\rm FL}=0.0249$ for $A_0=2.0$; the frequency of the slow oscillation in $\sigma_{xy}^T$ near the pulse peak coincides with the magnitude of the photoinduced gap. We note that such slow oscillations in the time-resolved Hall conductivity reflecting photon-dressed topological bands under pulse excitations have been reported also in graphene systems~\cite{Gavensky_PRB18}. For these values of $A_0$, the frequency gradually increases as $\tau$ increases toward the pulse peak, indicating a transient growth of the photoinduced gap. For $A_0=0.8$, the period of the slow oscillation is longer than those for $A_0=1.4$ and $2.0$, making precise estimations difficult. This means that the photoinduced gap is small. Indeed, for $A_0=0.8$, we have $2\pi/\Delta_{\rm FL}=167T$, which is longer than the pulse width ($\sigma_{\rm pu}=76T$); for smaller values of $A_0$, a longer pulse is needed to observe the slow oscillation in $\sigma_{xy}^T$ fully.

\section{Discussions and Summary}

We discuss the relevance of our results to experiments and the feasibility of the experiments. First, we mention the strength of the electric field of light considered in this study. In $\alpha$-(BEDT-TTF)$_2$I$_3$, the unit lengths in the $a$ and $b$ directions are close to $10$ \AA. From these values, we estimate that $A_0=1.4$ corresponds to $E^{\omega}=11.2$~MV/cm. It has been recently reported that pump-probe experiments using an intense pulse with $E^{\omega}$ exceeding 10~MV/cm have been successfully performed~\cite{Kawakami_NATP18,Kawakami_NATCM20}. These experimental reports support the feasibility of the proposed experiment using the pulse with $E^{\omega}$ as intense as 11.2~MV/cm.

Next, we discuss the electron-electron scattering effects which give rise to heating \cite{Schuler_PRX20}. To invoke the Floquet adiabatic picture under photoirradiation, the pulse width of the laser should be longer than the time scale of the photoinduced gap $\Delta_{\rm FL}$, which is of the order of 0.01~eV. Specifically, the oscillation period in the Hall conductivity shown in Figs. \ref{fig:fig3}(a) and \ref{fig:fig5} is $2\pi/\Delta_{\rm FL}=150\sim 400$ fs depending on the value of $E^{\omega}$. However, when the pulse is long, scattering effects due to electron-electron and electron-phonon interactions may become important. This point has been argued in recent studies of photoinduced dynamics in graphene \cite{Schuler_PRX20}. In graphene, photoirradiation with near-infrared light with frequency $\sim 1$ eV induces a large amount of photocarriers since the conduction and valence bands forming the Dirac cones have a wide bandwidth of about 15 eV. This photocarrier generation contributes to heating via electron-electron scattering \cite{Schuler_PRX20}. However, in $\alpha$-(BEDT-TTF)$_2$I$_3$, the four bands near the Fermi level lie in the energy range of 0.7 eV \cite{Kino_JPSJ06} which is comparable to the photon energy. Since these four bands are well separated by other upper and lower bands, this compound is a unique system where the off-resonance condition is realized by near-infrared light, which is in contrast to graphene. This makes the photocarrier generation ineffective and thus will result in considerable suppression of heating. In fact, for the setup of continuous-wave laser in Sec. III, the negligibly small amount of photocarriers is evident from the fact that we have $\alpha_{{\bm k},\nu}\sim 1$ for $\tau>0$ [see Fig. \ref{fig:fig2}(b)]. In addition, we have evaluated the occupied part of the excitation spectra $A_{\bm k}(\varepsilon,\tau_{\rm pr})$ in Figs. \ref{fig:fig4}(b)-\ref{fig:fig4}(d), which corresponds to the $G^<$ term in Eq. (\ref{eq:ak}), and have found that the electron occupation of the conduction band is negligibly small even during the photoexcitation process. In graphene, it has been argued that the anomalous Hall conductivity under circularly polarized light deviates from the quantized value expected from the Berry curvature of the Floquet Chern insulator, which has been ascribed to a large contribution from photocarriers \cite{Sato_PRB19}. In contrast, the Hall conductivity in Figs. \ref{fig:fig3}(a) and \ref{fig:fig5} exhibits a nearly quantized value. This indicates that it comes almost only from the Berry curvature of the photoinduced topological phase and thus $\alpha$-(BEDT-TTF)$_2$I$_3$ is a promising candidate for observing the Floquet Chern insulator through the quantized Hall conductivity. 

The slow oscillation with the period $\sim 2\pi/\Delta_{\rm FL}$ in the Hall conductivity [Figs. \ref{fig:fig3}(a) and \ref{fig:fig5}] would be damped by scattering effects. In organic compounds, the timescale of the electron-electron scattering is $\tau_e=2\pi/t\sim 40$ fs with $t$ being the typical transfer integral of 0.1 eV. However, it is expected that the electron-electron scattering does not severely hamper the slow oscillation in the Hall conductivity because the off-resonance condition offers a nearly coherent time evolution without the photocarrier generation.  On the other hand, the timescale of the electron-phonon scattering ($\tau_{\rm ph}$) that gives rise to dissipation would be one order of magnitude longer than that of $\tau_e$ and would be comparable to the pulse width. More specifically, we have used the pulse width $\sigma_{\rm pu}=76T\sim 400$ fs. Thus, if the electron-phonon scattering dominates the damping of the Hall conductivity, the amplitude of the slow oscillation in Fig. \ref{fig:fig5} will decrease to $e^{-\sigma_{\rm pu}/\tau_{\rm ph}}=37$ \% during the pulse.

The electron-electron interactions in $\alpha$-(BEDT-TTF)$_2$I$_3$ are manifested by the charge-order phenomenon that appears at ambient pressure. However, physical properties associated with the Dirac fermions have been well explained by weak coupling theories or even with the noninteracting model \cite{Katayama_JPSJ06,Katayama_EPJB08,Hirata_SCI17}. Moreover, it has been argued in graphene that the electron-electron and electron-phonon interactions have only little effect on the magnitude of the photoinduced gap as well as the Floquet band structure \cite{Schuler_PRX20}. These facts suggest the validity of our approach with the noninteracting model. However, heating and dissipation are inevitably present in photoinduced dynamics. Thus, it is important to examine their effects in order to clarify the experimental feasibility of our results. Also, under long laser pulses, possible sample damage may further reduce the experimental feasibility. In this sense, the creation of a large photoinduced gap with short pulses of strong electric field is favored in observing the dynamical gap formation associated with the Floquet topological insulator phase in $\alpha$-(BEDT-TTF)$_2$I$_3$.

In summary, we have investigated the real-time dynamics of the photoinduced topological state in the organic conductor $\alpha$-(BEDT-TTF)$_2$I$_3$ with a pair of tilted Dirac cones in its energy-band structure. We have solved the time-dependent Schr\"odinger equation numerically for the tight-binding model of $\alpha$-(BEDT-TTF)$_2$I$_3$ coupling with an ac electric field of circularly polarized light. For the continuous-wave excitations, time profiles of the Chern number and the Hall conductivity demonstrate the appearance of the Floquet topological insulator phase that was predicted using the Floquet theory~\cite{Kitayama_PRR20}. We have shown that the Hall conductivity has a slow oscillation component for which the frequency coincides with the photoinduced direct gap at the Dirac point. For the pulse excitations, we have calculated transient excitation spectra, by which the formation of the Floquet bands with the photoinduced gap is elucidated in a time-resolved manner. We have shown that the slow oscillation component of the Hall conductivity exhibits the signature associated with dynamical growth of the gap during the pulse irradiation.

\begin{acknowledgments}
This work was partly supported by JSPS KAKENHI (Grants No. 17H02924, No. 16H06345, No. 19H00864, No. 19K21858, No. 19K23427, No. 20K03841, and No. 20H00337) and Waseda University Grant for Special Research Projects (Projects No. 2019C-253 and No. 2020C-269). 
\end{acknowledgments}

\end{document}